\documentclass[aps,pra,superscriptaddress,amssymb,showpacs]{revtex4}
\usepackage{setstack}
\usepackage{graphicx}

\begin{document}

\title { Pulsed "three-photon" light}

\author{T.~V.~Gevorgyan}
\affiliation{Institute for Physical Researches, National Academy
of Sciences,\\Ashtarak-2, 0203, Ashtarak, Armenia}
\email[]{t\_gevorgyan@ysu.am}
\author{G.~Yu.~Kryuchkyan}
\email[]{kryuchkyan@ysu.am}
\affiliation{Yerevan State University, Alex Manoogian  1, 0025,
Yerevan, Armenia}\affiliation{Institute for Physical Researches,
National Academy of Sciences,\\Ashtarak-2, 0203, Ashtarak,
Armenia}

\begin{abstract}

Generating multi-photon entangled states is a primary task for
 applications of
quantum information processing.  We investigate production of photon-triplet
in a regime of light amplification in second-order nonlinear media under action of a
 pulsed laser beam. For this goal the process of cascaded three-photon
splitting in an optical cavity  driven by a sequence of laser
pulses with Gaussian time-dependent envelopes is investigated. Considering production of
photon-triplet for short-time regime and in the cascaded three-wave collinear configuration
  we shortly analyze
 preparation of polarization-non-product  states  looking further
applications of these results in the cascaded optical parametric oscillator.
 It is also demonststed the nonclassical characteritics of the photon-triplet
 in phase-space on the base of the Wigner function. Calculating the
normalized third-order correlation functions below-and at the
generation threshold of cascaded optical parametric oscillator, we demonstrate that in the pulsed regime,
depending on the duration of pulses and the time-interval
separations between them, the degree of three-photon-number
correlation essentially exceed the analogous one for the case of
continuous pumping.

\end{abstract}
\pacs{03.65.Ud, 03.65.Yz}

\maketitle

\section{Introduction}

Multiphoton entangled states have attracted a great interest in
probing the foundations of quantum theory and constitute a
powerful quantum resource with promising potential for various
applications in quantum information technologies. Recently,
experimental efforts in the direct production of multiphoton joint
states, particularly, three- or four-photon states  have paved a
new stage for the study of multipartite entanglement \cite{Hamel}.
Indeed, the simultaneous generation of three photons is at the
origin of intrinsic three-particle quantum properties such as
Greenberger-Horne-Zeilinger (GHZ) -class and W-class quantum
entanglement \cite{H2} - \cite{H5}. Up to now, several physical
systems have been proposed for the generation of photon triplet
including third-order nonlinear medium \cite{H6} and cascaded
spontaneous parametric down-conversion (PDC) \cite{H7,H8}.
Experimentally, three-photon down-conversion was studied in
third-order nonlinear media \cite{H9} - \cite{H11} and also by
using cascaded second-order nonlinear parametric processes
\cite{Hl2}. Direct generation of photon triplets using cascaded
photon-pairs has been demonstrated in periodically poled lithium
niobate crystals \cite{Hamel}. The distinction of three-photon GHZ
and W states entangled in time and space has been also reported
\cite{Hl3, Hl4}. It was also shown that intracavity three-photon
down-conversion can be effectively realized in cascaded optical
parametric oscillator (OPO) \cite{Hl5}. This scheme that involves
cascading second-order nonlinearities is based on the parametric
processes of splitting and summing in which the frequencies
between the pump and two subharmonics frequencies are in the ratio
of $3: 2: 1$. Experimental realization of cascaded OPO by using
the dual-grid method of quasi-phase matching (QPM) has been done in
Ref. \cite{Hl6}. Most recently, joint quantum states of
three-photons with arbitrary spectral characteristics have been
studied on the base of optical superlattises \cite{Hl7} for the
cascaded configuration proposed in \cite{Hl5}. Two cascaded
configurations have been considered in \cite{Hl7} that lead to
production of spontaneous photon triplet in cascaded PDC and
generation of high intensity mode due to cascaded three-photon
splitting in optical cavity.

In this paper we continue the investigation of cascaded
three-photon splitting in an optical cavity following the paper
\cite{Hl7}. Our goal is twofold. In one part of the present paper
we extend our previous results regarding three-photon splitting in
optical cavity for an experimentally available scheme that is a
cascaded parametric oscillator pumped by a sequence of Gaussian
laser pulses (see, Sects. II and III). The other part of the paper is devoted to studies of
quantum properties of "three-photon" mode. We  discuss
 preparation of non-product states that are superposition of
 three-photon polarization states, however, without any consideration of cavity effects
(see, Sec. IV).  We also calculate the Wigner function of the subharmonic,
 i.e. "three-photon mode" showing the negativity in phase-space (see, Sec. V).
Our analysis includes
calculation of third-order correlation function of
photon numbers for various operational regimes of pulsed OPO(see, Sec. VI).
 It is known, that it is possible to control the behavior of quantum
dissipative system by a train of pulses. In
this paper, we use this approach for suppression of dissipation
and cavity induced feedback in cascaded OPO that leads to
increasing the level of three-photon-number correlation.

\section{Periodically pulsed cascaded OPO: Generation threshold}\label{SecPD}

In this section we briefly describe the cascaded optical parametric oscillator (OPO) with a triply
resonant optical ring cavity driven by a sequence of laser pulses.
The semiclassical and quantum theories of this device for the
monochromatic pump field were developed in Refs. \cite{Hl5, Hl7}
and here we only add some important details regarding cascaded OPO
under laser pulses with Gaussian time-dependent envelopes. This
cascaded configuration involves the fundamental mode driven by an
external pump field at the central frequency $\omega_{0}$ with an
amplitude $E_{L}$ and two subharmonic modes at the frequencies
$\omega_{1}=\frac{\omega_{0}}{3}$ and
$\omega_{2}=\frac{2\omega_{0}}{3}$. Due to intracavity parametric
type-I three-wave interactions pump field is converted to the
subharmonics throughout two cascaded processes:
$\omega_{0}\rightarrow \omega_{1}+\omega_{2}$ and
$\omega_{2}\rightarrow \omega_{1}+\omega_{1}$. Subharmonic modes
have the same plane polarizations and are all propagating in the
same direction. The pump field consists from the sequence of
Gaussian laser pulses with the amplitude

\begin{equation}
E(t, z)=E_{L}f(t)e^{-i(\omega_{0}t -k_{L}z)},
\end{equation}

\begin{equation}
f(t)=\sum_{n=0}^{\infty}e^{-(t-t_{0}-n\tau)^{2} /
T^{2}},\label{E2}
\end{equation}
where $ T $ is the duration of pulses that are separated by time
intervals $\tau$.

 The cascaded OPO is dissipative, because the
modes suffer from losses due to partially transmission of light
through the mirrors of the cavity. We consider below the case of
high cavity losses for the pump mode $(\gamma_{0} \gg \gamma,
\gamma_{1}=\gamma_{2}=\gamma)$, when the pump mode is eliminated
adiabatically in non-depletion approximation. In this case, the
effective interaction Hamiltonian in the rotating wave
approximation reads as
\begin{eqnarray}
H_{int}=i\hbar\chi_{1} E_{0}f(t)\left(a_{1}^{+} a_{2}^{+}-a_{1}
a_{2} \right) +i\hbar\chi_{2}\left(a_{1}^{+2}
a_{2}-a_{1}^{2}a_{2}^{+}\right),~~~~~\label{Hamint}
\end{eqnarray}
where $E_{0}=E_{L} / \gamma_{0}$, $a_{i} $ (i=1,2) are the
operators of the modes at the frequencies
$\omega_{1}=\frac{\omega_{0}}{3}$ and
$\omega_{2}=\frac{2\omega_{0}}{3}$ and the coupling constants
between the modes are expressed through the Fourier spectra of the
second-order susceptibilities of nonlinear crystals of the length
$L$
\begin{eqnarray}
\chi_{1}=\int^{L}_{0}dz\chi^{\left(2\right)}(z)e^{i\Delta
k_{1}(z)z},\label{zt}\\
\chi_{2}=\int^{L}_{0}dz\chi^{\left(2\right)}(z)e^{i\Delta
k_{2}(z)z}.\label{qsi}
\end{eqnarray}
We assume collinear, one-dimensional on $ z$
quasi-phase-matching with the phase mismatch vectors
\begin{equation}
\Delta k_{1}(z)=k_{L}(\omega_{0}, z)-k_{1}(\omega_{1},
z)-k_{2}(\omega_{2}, z),
\end{equation}
\begin{equation}
\Delta k_{2}(z)=k_{2}(\omega_{2}, z)-2k_{1}(\omega_{1}, z)
\end{equation}  analyzed in the details \cite{Hl7}.

In this regime, the stochastic equations of motion for the complex
c-number variables $\alpha_{1,2}$ and $\beta_{1,2}$ corresponding
to the operators $a_{1,2}$ and $a_{1,2}^{+}$, have the following
form
\begin{equation}
\frac{d \alpha_{1}}{d t}=-\gamma_{1}\alpha_{1}+
E_{0}f(t)\chi_{1}\beta_{2} + 2\chi_{2}\alpha_{2}\beta_{1} +
W_{\alpha_{1}}(t),\label{dalfa1}
\end{equation}
\begin{equation}
\frac{d \beta_{1}}{dt}=-\gamma_{1}\beta_{1}+
E_{0}f(t)\chi_{1}\alpha_{2} +
2\chi_{2}\beta_{2}\alpha_{1}+W_{\beta_{1}}(t).\label{dalfa2}
\end{equation}
The equations for $\alpha_{2},\beta_{2}$ are obtained from (8),
(9) by exchanging the subscripts $(1) \rightleftharpoons (2)$. Our
derivation is based on the Ito stochastic calculus, and the
nonzero stochastic correlators are:
\begin{equation}
\langle W_{\alpha_{1}}(t)W_{\alpha_{2}}(t^{'}) \rangle = \chi_{1}
\frac{E_{L}f(t)}{\gamma_{0}}\delta(t-t^{'}), \label{w1}
\end{equation}

\begin{equation}
\langle W_{\alpha_{1}}(t)W_{\alpha_{1}}(t^{'}) \rangle =
2\chi_{2}\alpha_{2}\delta(t-t^{'}), \label{w2}
\end{equation}

\begin{equation}
\langle W_{\beta_{1}}(t)W_{\beta_{1}}(t^{'}) \rangle =
2\chi_{2}\beta_{2}\delta(t-t^{'}), \label{w3}
\end{equation}

\begin{equation}
\langle W_{\alpha_{2}}(t)W_{\alpha_{2}}(t^{'}) \rangle =
2\chi_{2}\alpha_{1}\delta(t-t^{'}). \label{w4}
\end{equation}
The Eqs. (\ref{dalfa1}), (\ref{dalfa2}) and the correlation
functions modifies the analogous ones derived for OPO with
monochromatic pumping \cite{Hl5} on the case of non-stationary
pump field.

In accordance with the cited paper, for the monochromatic driven
cascaded OPO, zero-amplitude solutions $\alpha_{1}=\alpha_{2}=0$
of Eqs. (\ref{dalfa1}), (\ref{dalfa2}) with $f(t)=1$ are stable,
if $E_{L} < E_{th} =
\frac{\gamma_{0}}{\chi_{1}}\sqrt{\gamma_{1}\gamma_{2}}$, while
steady-state photon numbers $n_{1}$ and $n_{2}$ display
histeresis-cycle behavior in a small domain $\frac{2\sqrt{2}}{3}<
E_{L}/E_{th}<1$. Thus, remarkable feature of OPO under
monochromatic pump is comparatively low generation threshold that depends from the second-order susceptibility in
comparison with the scheme of direct intracavity three-photon
down-conversion, where the pump power threshold is determined by
the third-order susceptibility \cite{H9}.

Below, we derive the threshold value for OPO driven by trains of
Gaussian pulses. The analysis of stochastic equations shows that
similar to the standard OPO with monochromatic pump field amplitude,
the periodically pulsed OPO also exhibits threshold behavior,
which is easily described through the period averaged pump field
amplitude $\overline{f(t)}=\frac{1}{\tau}\int_{0}^{\tau}f(t)dt$.
We demonstrate this statement analyzing the stability of
zero-amplitude solutions of $\alpha_{1}=\alpha_{2}=0$  of Eqs.
(\ref{dalfa1}), (\ref{dalfa2}) for both modes below threshold and
for the case when the decay rates of subharmonics are equal one to the order, $\gamma_{1}=\gamma_{2}=\gamma$. To check the stability we turn
to the linearized on the small deviations $\delta \alpha_{i},
\delta \beta_{i}$  the equations in the semicalssical approach without noise terms.
These equations can be rewriten in the following
form:
\begin{equation}
\frac{d}{dt}\delta X_{\pm}=(-\gamma \pm
\frac{E_{L}}{\gamma_{0}}\chi_{1} f(t))\delta X_{\pm},
\label{dalfa12}
\end{equation}
\begin{equation}
\frac{d}{dt}\delta Y_{\pm}=(-\gamma \pm
\frac{E_{L}}{\gamma_{0}}\chi_{1} f(t))\delta Y_{\pm},
\label{dalfa13}
\end{equation}
through the quadrature field variables defined as $\delta
\alpha_{\pm}=\delta \alpha_{1} \pm \delta \alpha_{2}$ and $\delta
X_{\pm}=\frac{1}{2}(\delta \alpha_{\pm} + \delta
\alpha^{*}_{\pm}), \delta Y_{\pm}=\frac{1}{2i}(\delta \alpha_{\pm}
- \delta \alpha^{*}_{\pm})$. In these variables the time evolution
has the simple form

\begin{equation}
\delta X_{\pm}(t)=\Lambda_{\pm}(t,t_{0}) \delta X_{\pm}(t_{0}),
\label{E16}
\end{equation}

\begin{equation}
\delta Y_{\pm}(t)=\Lambda_{\mp}(t,t_{0}) \delta Y_{\pm}(t_{0}),
\label{E17}
\end{equation}

\begin{equation}
\Lambda_{\pm}(t,t_{0})=exp(\pm
\frac{E_{L}\chi_{1}}{\gamma_{0}}\int^{t}_{0}f(t^{'})dt^{'}-\gamma(t-t_{0})).
\label{E18}
\end{equation}

Analyzing semiclassical equations and operational regimes we
choose the switching time in infinity, i.e. $t_{0}\rightarrow
-\infty$, and add in (\ref{E2})  terms with negative  $n$. In this
case, the function $f(t)$ is periodic on time $f(t+\tau)=f(t)$ and
the analysis is simplified. Since the function $f(t)$ is periodic
on time, we can obtain the general formula
$\int^{t}_{t_{0}}f(t)dt=\overline{f(t)}(t-t_{0}) +F(t)- F(t_{0})$,
where $F(t)$ is a periodic function, $F(t+\tau)=F(t)$. Therefore,
we see from (\ref{E16}), (\ref{E17}), (\ref{E18}) that the
solution $\alpha_{i}=0$ below-threshold is stable if $E_{L}<
\frac{\gamma_{0}\gamma}{\chi_{1}}\overline{f(t)}$. It is easy to
check also that due to noted periodicity of the amplitude the following formula
takes place (see, for example \cite{HADAM})
\begin{equation}
\int_{0}^{\tau}dt \sum_{n=-\infty}^{\infty}e^{-(t-n\tau)^{2} /
T^{2}}=\int_{-\infty}^{\infty}dte^{-t^{2} / T^{2}}.\label{E19}
\end{equation}
This formula allows us to calculate the averaged value of the
amplitude $f(t)$. On the whole, we arrive to the result that for
the case of Gaussian pulses above threshold regime is realized if
\begin{equation}
E_{L}\geq
\overline{E_{th}}=\frac{\tau}{T\sqrt{\pi}}E_{th}=\frac{\tau}{T\sqrt{\pi}}
\frac{\gamma_{0} \gamma}{\chi_{1}}.
\end{equation}
The important peculiarity of the system proposed is that the
threshold value $\overline{E_{th}}$ depends on the coupling
constant $\chi_{1}$ which is related to the second-order
susceptibility as well as depends on the characteristics of laser
pulses.

\section{Numerical simulation of dissipation and decoherence}\label{SecPD}

The cascaded OPO is dissipative, because the modes suffer
from losses due to partial transmission of light through the
mirrors of the cavity and due to quantum fluctuations.  We analyse dissipative and decoherence effects
on the base of
master equation for the density operator of the cavity modes in
the Lindblad form
\begin{equation}
\frac{\partial\rho}{\partial
t}=\frac{1}{i\hbar}[H_{int},\rho]+\sum_{i=1,2}{\gamma_{i}\left(2a_{i}\rho
a^{+}_{i}-a^{+}_{i}a_{i}\rho-\rho
a^{+}_{i}a_{i}\right)}.\label{dRo}
\end{equation}
We calculate the quantities of interest (the photon number distributions, Wigner functions, etc.)
mainly for the subharmonic mode (1)
by using the reduced density operator $\rho_{1}(t)$ which is
constructed from the density operator $\rho(t)$ of both modes by
tracing over the mode (2), $\rho_{1}(t)=Tr_{2}(\rho)$. We analyze
the master equation numerically using quantum state diffusion
method (QSD) \cite{H18},  \cite{claster}. According to this method, the reduced
density operator is calculated as the ensemble mean
\begin{equation}
\rho(t)= M(|\psi_{\xi}(t)\rangle\langle\psi_{\xi}(t)|)= \lim_
{N\rightarrow\infty}\frac{1}{N}\sum_{\xi}^{N}|\psi_{\xi}(t)\rangle\langle\psi_{\xi}(t)|
\end{equation}
over the stochastic pure states $|\psi_{\xi}(t)\rangle$ describing
evolution along a quantum trajectory. The stochastic equation for
the state $|\psi_{\xi}(t)\rangle$ involves Hamiltonian described
by Eq. (\ref{Hamint}) and the Lindblad operators described by noise
terms in the master equation (\ref{dRo}). We calculate the density
operator using an expansion of the state vector
$|\psi_{\xi}\rangle$ in a truncated basis of Fock's number states
of modes of the subharmonics (1) and (2)
\begin{equation}
|\psi_{\xi}(t)\rangle= \sum_{n}a_{n_{1},
n_{2}}^{\xi}(t)|n_{1}\rangle_{1}|n_{2}\rangle_{2}.
\end{equation}

Details of analogous calculations for an anharmonic oscillator in
time-modulated field can be found in \cite{H19}). The numerical
simulations are performed in the truncated Fock basis of the
subharmonic modes that are limited by 500 photons. This
approximation is valid for the case of strong nonlinear couplings
$\chi_{1}$ and $\chi_{2}$ with respect to the  dissipation
parameters.

\section{Production of  polarization, non-product states of photon triplet in collinear configuration}\label{SecW}

Three-photon correlations allow the creation of tripartite
entangled states such as the GHZ state. For cascaded SPDC in
noncollinear configuration spatially-polarization GHZ states has
been considered in \cite{Hl7}. Below we apply the results obtained
for preparation of three-photon polarization states in collinear
configuration of interacting waves. Recently,  a simple but highly efficient
 source of polarization-entangled photon pairs at nondegenerate
wavelengths and in collinear configuration has been demonstrated \cite{HAPL}.
We consider the production of polarization-entangled photon triplet.
It is possible for
the case when cascaded processes involve polarized photons.

Thus,
we modify the above results considering three-wave interaction
with the indexes of polarization states. Looking further
applications of above results for intracavity three-photon
down-conversion in collinear configuration of cascading processes
we concentrate on consideration of non-product states that are entangled
only on polarization degree of freedom but not on spatial
variables. Thus, including into consideration the polarization
states of the photons we assume that the type-II process
$\omega_{0}\rightarrow\omega_{1}+\omega_{2}$ create the pair of
photons with $a$ vertical $V, (a_{1}(V), (a_{2}(V)) $ and
horizontal $H, (a_{1}(H),(a_{2}(H)) $ polarizations in collinear
configuration. If the pump field is oriented at $45^{\circ}$ to
the horizontal and vertical axes two processes
$\omega_{0}\rightarrow\omega_{1}(V)+\omega_{2}(H)$ and
$\omega_{0}\rightarrow\omega_{1}(H)+\omega_{2}(V)$ take place in
the first crystal. The next process is considered as the type-I parametric process.
In type-I conversion, photon pairs are created with the same polarization state, but
ortogonal to the input mode. Therefore, the second, type-I crystal is arranged in the
manner that the following process:
$\omega_{2}(H)\rightarrow\omega_{1}(V)+\omega_{1}(V)$ and
$\omega_{2}(V)\rightarrow\omega_{1}(H)+\omega_{1}(H)$ should be
realized.

For simplicity, we restrict our attention considering
frequency-uncorrelated three-photon states and assume that the
process under photons with $(V)$ and $(H)$ polarizations are
described by the equal coupling constants. Thus, we assume that
photon pairs in three-wave processes :
$\omega_{0}\rightarrow\omega_{1}+\omega_{2}$,
$\omega_{2}\rightarrow\omega^{'}_{1}+\omega^{"}_{1}$ have
correlations on the polarization, but not on the spectral lines.
In analogy with Eq.(3), we model the sum of the corresponding
parametric interactions by the following effective Hamiltonian
\begin{eqnarray}
H=H_{1}+H_{2},\label{ankH}~~~~~~~~~~~~~~~~\\
H_{1}=i\hbar\chi
E_{0}f(t)(a^{+}_{1}(V)a^{+}_{2}(H)+a^{+}_{1}(H)a^{+}_{2}(V))+h.c.,\\
H_{2}=i\hbar
k[(a_{2}(V)(a^{+}_{1}(H))^{2}+a_{2}(H)(a^{+}_{1}(V))^{2})]+h.c..\label{HGHZ}
\end{eqnarray}
Here, $a_{1}(V)$ and $a_{2}(V)$ are the annihilation operators of
modes (1) and (2) at vertical polarizations, while the operators
$a_{1}(H), a_{2}(H)$ corresponds to the horizontal polarized
photons of the frequencies $\omega_{1}$ and $\omega_{2}$,
respectively.

We focus on analysing the generation of non-product states  for one-passing configuration of
cascaded parametric spontaneous processes
without consideration of cavity dynamics and feedback effects. This approach
is valid for short interaction time intervals much shorter
than the characteristics relaxation time. In this case
time-evolution of the vector state of the system is described by
the second-order term of the perturbation theory. Choosing the initial
state as a vacuum state
$|\psi_{in}\rangle=|0\rangle_{a_{1}(H)}|0\rangle_{a_{2}(H)}|0\rangle_{a_{1}(V)}|0\rangle_{a_{2}(V)}$
for all modes we derive the final state during time evaluation in
the following form

\begin{eqnarray}
|\psi(t)\rangle=\left(-\frac{i}{\hbar}\right)^{2}t^{2}H_{2}H_{1}|\psi\rangle_{in}= \nonumber ~\\
\bar{\chi}\bar{k}E_{0} (| V \rangle | V \rangle
| V \rangle + | H \rangle| H \rangle| H
\rangle)|0\rangle_{a_{2}(H)}|0\rangle_{a_{2}(V)},\label{PsGHZ}
\end{eqnarray}
where $\bar{\chi}=\chi t$  and
$\bar{k}=kt$ are the coupling constants and the states
$|V\rangle=a^{+}_{1}(V)|0\rangle_{a_{1(V)}}$,
$|H\rangle=a^{+}_{1}(H)|0\rangle_{a_{1(H)}}$ present the vertical
and horizontal polarization states of photons at the frequency
$\omega_{1}=\frac{\omega_{0}}{3}$. Thus, we demonstrate that in
this collinear, one-dimensional cascaded scheme triple photons can
constitutes the polarization entangled (non-product) states of light.
It should be noted that under cavity feedback effects the non-product quantum state
cannot be described by this simple expression and we should include the higher-order terms
of the perturbation expansion into consideration.

\section{Photon triplet in phase-space: Wigner functions and photon number distributions in the pulsed regime}

Quantum interference signature of three-photon
states in phase-space  has been demonstrated for the direct three-photon down-conversion in
third-order nonlinear medium \cite{H6, H9} as well as in the cascaded scheme \cite{Hl7}
for the case of monochromatic pumping. We demonstrate now this
effect for the pulsed regime of cascaded OPO.  We illustrate
these effects numerically on the base of the master equation,
however, in the regimes when the dissipation in the cavity is
unessential and the dynamic of modes is almost unitary. For the
cavity configuration presented, the validity of such approximation
is guaranteed by consideration of short interaction time for which the duration
of pulses are much shorter than the characteristics relaxation
time, $1/(\chi_{1}E_{0}),1/\chi_{2}<<t<<1/\gamma_{1, 2}$ ,  provided that the nonlinear coupling constants
exceed the dumping rates for the modes.

 Below, we present the results on the photon-number distributions and the Wigner functions
 for three-photon mode. The photon number
distribution for $\omega_{1}=\frac{\omega_{0}}{3}$  mode is calculated as
the diagonal element $P_{1}(n)=\langle n|\rho| n \rangle$ of photonic Fock states while calculations of the Wigner function for the
this mode are performed by using its standard formula in a Fock
space:

\begin{eqnarray}
W_{1}(\rho, \theta)=\sum_{m,n}\rho_{1,mn}W_{mn}(\rho, \theta).
\end{eqnarray}
Here, $\rho, \theta$  are the polar coordinates in the complex
phase space which is determined by the position and the momentum
of  quadratures $x = (a+a^{+})/ \sqrt{2},  y = (a-a^{+})/ \sqrt{2i}$,
respectively, while the coefficients $W_{mn}(\rho, \theta) $  are the Fourier
transforms of the matrix elements of the Wigner characteristic
function.

\begin{figure*}
\begin{center}
\includegraphics[width=15cm]{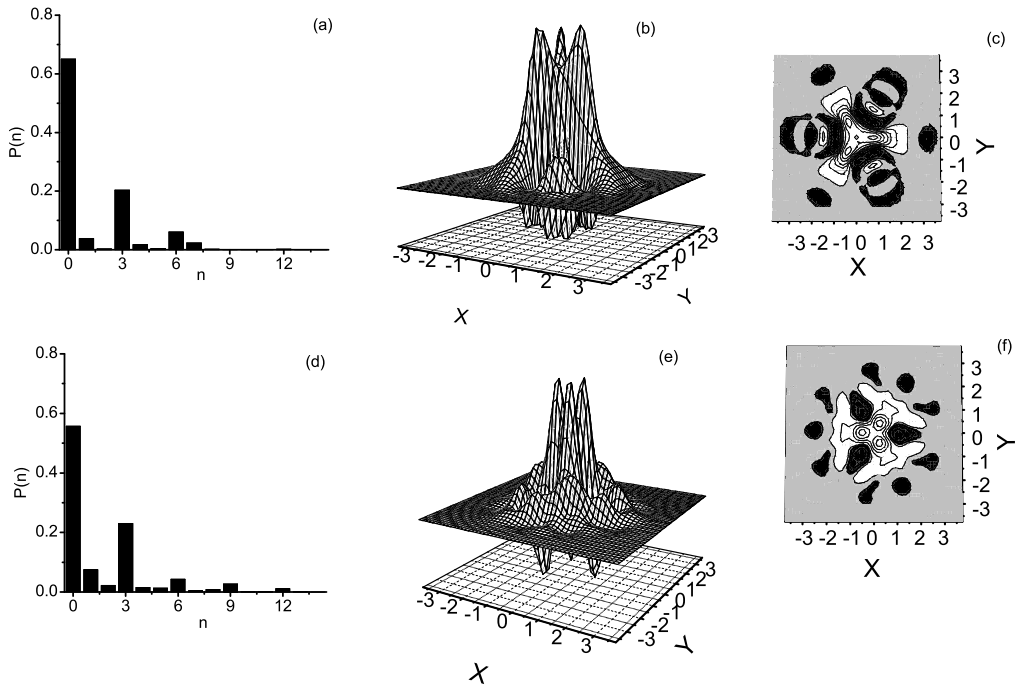}
\end{center}
\caption{The photon number distributions (a, d), the Wigner functions
(b, e) and its contour plots (c, f) for $\omega_{1}$-mode and for
different time intervals: $t= t_{0} + 6.2*10^{-3} \gamma^{-1} $
(a, b, c); $t= t_{0} + 6.2*10^{-2} \gamma^{-1}$ (d, e, f). The other
parameters are: $\chi_{1}/\gamma_{1}=200$,
$\chi_{2}/\gamma_{1}=100$, $\gamma_{2} = \gamma_{1}=\gamma, T \gamma
 = 1*10^{-3}, \tau \gamma = 6*10^{-3}$ .}\label{9W}
\end{figure*}

Three-photon structure of the mode is shown on the photon-number distributions in Figs. 1 (a, d).
As we see, for short time-intervals, the most probable
values of photon numbers are separated by three photons. In Figs.1 (b, e) we present
results on the Wigner functions that clearly exhibit phase-space
quantum interfringes. These results describe the
case of cascaded OPO under two consecutive pulses with the
duration $T \gamma = 1*10^{-3}$ separated by the time-interval $
\tau \gamma = 6*10^{-3}$.  Fig.1(b) shows the Wigner function evolved
for a time interval $t= t_{0} + 6.2*10^{-3} \gamma^{-1} $ that corresponds to maximal
photon number $n_{max}$ of the first pulse; Fig.1 (e) shows the
Wigner function at $t= t_{0} + 6.2*10^{-2} \gamma^{-1}  $ corresponding to $n_{max} $ of
the second pulse.
The Wigner function show three phase components with an
interference pattern in the regions between them. We show
the regions of quantum interference in the contour plots (see,
Figs.1 (c, f)) depicting negative regions of the interference terms
in black. Note that threefold symmetry of the Wigner function
and interference pattern has been demonstrated for the direct
three-photon down-conversion in $\chi^{(3)}$ media  \cite{H6, H9}. However,
we note that the results presented here for the pulsed cascaded
configuration are also different in details from the analogous
calculation of the Wigner function for the case of monochromatic pump field  \cite{Hl7}.

\section{Photon-number correlation in the pulsed regime}

The experimental verification of time-dependent correlation between photons
in triplet has been demonstrated for one-passing configuration of
cascaded SPDC \cite{Hamel}. Considering production of photon
triplet in a cavity, it seems that the correlation between photons
can be evidently displayed for short interaction time intervals
much shorter than the  relaxation time.
Nevertheless, the three-photon number correlation exceeding the
coherent level, (that means the normalized third-order correlation
function $g^{(3)}>1$),
 has been demonstrated for the cascaded OPO
driven by monochromatic pumping, in over transient regime  for modes generated
 below the threshold \cite{Hl7}. This
effect decreases if the system moves to the range of the
generation threshold. At the threshold, the typical value for the
normalized third-order correlation function for zero delay-time,
 $g^{(3)} = 1.2$  has been
obtained.

Note, that effects of  two-photon correlations for ordinary OPO and NOPO have  been a subject of intense research efforts for last years. These problems have been investigated in details in a  linear approximation on quantum fluctuations as well as  in the framework of  exact quantum theory of intracavity  parametric generation with allowance for quantum noise of arbitrary intensity.  In this way, the critical behavior of the second-order correlation functions
which describe photon correlation effects has been found analytically on the base of the Fokker-Planck equation for the density matrix in the threshold region of generation \cite{critic}.

\begin{figure}[h]
\includegraphics[width=15cm]{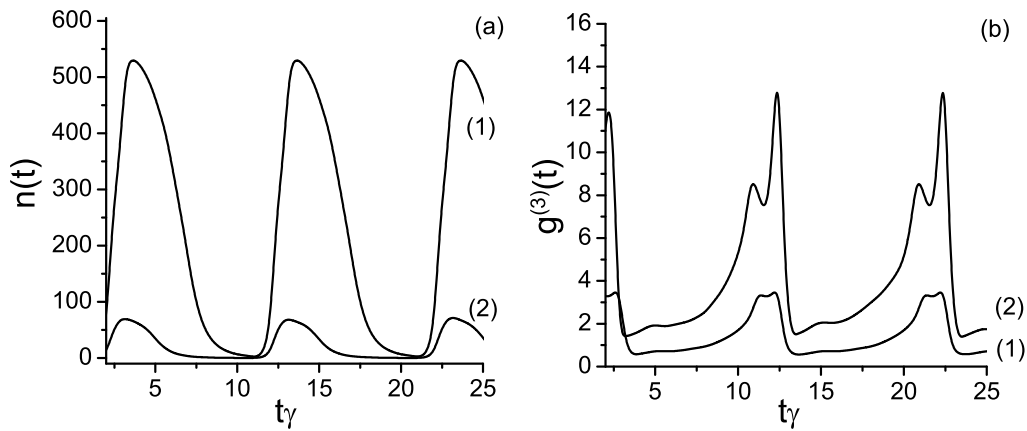}\\
\caption{The mean photon number (a) and the third-order correlation
function (b) versus $t\gamma$ at the threshold
$E_{L}/\overline{E_{th}}=1$, curves (1) and below the threshold
$E_{l}/ \overline{E_{th}}=0.5$, curves (2). The parameters are:
$\chi_{1}/\gamma=0.2$, $\chi_{2}/\gamma=0.1$,
$\gamma_{2}=\gamma_{1}=\gamma$, $\tau \gamma = 10$,
 $T \gamma =1$.} \label {ff1}
\end{figure}

In this Secion, we demonstrate the new regimes of strong
three-photon correlation for the pulsed cascaded OPO. We
concentrate on the numerical simulation of both the mean photon number of
subharmonics as well as the third-order correlation function.
Let us now discuss photon-number correlation in the time domain,
considering output twin light beams from the pulsed cascaded OPO
on the base of normalized third-order photon-numbers correlation
function $g^{(3)}$ for the mode (1)
\begin{equation}
g^{(3)}(t)=Tr( a_{1}^{+3} a_{1}^{3} \rho_{1}(t))  / n^{3}(t).
\end{equation}
Here, $n(t)=Tr( a_{1}^{+} a_{1} \rho_{1}(t))$ is the mean photon
number. Considering three-photon number correlation for the
intensive cavity mode in the presence of dissipation and cavity
induced feedback, we control quantum dynamics of dissipative
systems by the train of pulses. Here, we use this approach for
controlling quantum statistics of mode, particularly for
increasing the level of three-photon-number correlation. We
analyze the cases in which $T\gamma \leq 1$ and $\tau\gamma \geq
1$, for over transient time-intervals, $t \gg \gamma^{-1}$,
considering the operational regimes below-and  at the generation
threshold. Typical results for the mean photon numbers  and the
correlation function $g^{(3)}$ for two different parameters of the
Gaussian pulses :  $ T=\gamma ^{-1}, \tau =10\gamma ^{-1}$ and $
T= \gamma ^{-1}, \tau =5\gamma ^{-1}$ are presented in Figs.
(\ref{ff1}) and (\ref{ff3}).
\begin{figure}[h]
\includegraphics[width=15cm]{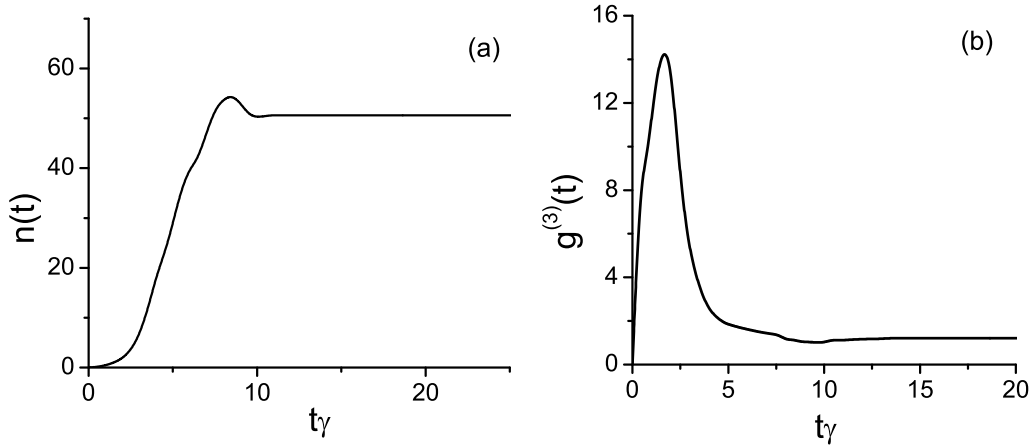}\\
\caption{Mean photon number (a) and the correlation function (b)
 for continuously pumped OPO at the threshold
$E_{L}/E_{th}=1$. The parameters are: $\chi_{1}/\gamma=0.2$,
$\chi_{2}/\gamma=0.1$, $\gamma_{2} =
\gamma_{1}=\gamma$.}\label{ff2}
\end{figure}
As we see from these figures, the time-dependence of these
quantities repeat the periodicity of the pump laser over transient
time intervals.
 We also conclude that
the maxima of the three-photon correlation are realized for the
definite time intervals for which the mean photon number of mode
(1) is in the ranges between its maxima and minima. As shown our
calculations, such strong photon correlations take place in
nonstationary regime of cascaded OPO when duration of pulses is
close to a characteristic dissipation time. In Figs. (\ref{ff1})
we analyze the mean photon number (Fig. \ref{ff1}(a)) and the
correlation function (Fig.\ref{ff1}(b)) in the operational regimes
of OPO below- and at the generation threshold. In the regime below the
generation threshold, at $E/ \overline{E_{th}}=0.5$ the
correlation function display two-peak structure (see, curve (2)).
The lower peak corresponds to the time intervals for which the
mean photon number is between its minimal and maximal values. More
correctly, for the period trains of intervals $t=t_{0}+m11.6
\gamma^{-1}, (m=1, 2, ...)$, the mean photon number $n=50$ and the
correlation function $g^{(3)}=8.9$. The other peaks correspond to
the minimal values of the mean photon number. At the threshold
(see, curves (1)) the effect of photon correlation is decreased,
although the level of correlation exceeds the coherent level,
$g^{(3)}>1$, particularly, we get $g^{(3)}=3.4$ for the mean
photon number n=251. Thus, we found a remarkable result that the
degree of three-photon number correlation for the pulsed regime of
OPO surpass the analogous result for OPO with continuous pumping
for the same mean photon numbers. Indeed, this conclusion is
illustrated in Figs. (\ref{ff1}) and (\ref{ff2}), where the
comparison of the results on pulsed regime with the calculations
based on the Hamiltonian (\ref{Hamint}) with $f=1$ is done. Note,
that the ideal limit of continuous pumping is realized if
$T\rightarrow 0, \tau \rightarrow \infty$ for the case of infinity
numbers of pulses. We present on the Figs. (\ref{ff2}) the results
for OPO with continuous pumping at the threshold $E_{L}=E_{th}$,
in which the mean photon number $n=52$ in the steady state regime
(curve (2), Fig. \ref{ff1}(a)) approximately equals to the maximal
value of the photon number in the pulsed regime. However, as we
see, the level of the maximal correlation, $g^{(3)}=9$ in this
case exceeds the analogous one for the case of continuous pumping,
$g^{(3)}=1.2$.

It is natural to explain such improvement of three-photon
correlation by control the behavior of a quantum system by an
external time-dependent force. The presence of these effects in
the cascaded OPO, particularly, can been seen from the
noise-correlation functions. Indeed, the equation (\ref{w1})
describes a multiplicative noise-term, where the level of noise is
determined by the amplitude of pulsed driving field $E_{L}f(t)$
leading to the control of dissipation. In this spirit, we
emphasize that the idea of controlling the dynamics of a quantum
system in the presence of dissipation and decoherence by an
external periodic driving was exploited by many authors (see, for
example, \cite{H20} and the references therein). In one of the
standard techniques control of the optical quantum system is
achieved through the application of suitable tailored,
synchronized laser pulses \cite{H21}. In this way, it is
interesting to analyze the three-photon correlation function in
dependence of the time-separation between pulses, i.e. for the
other parameters of driving pulses in additional to the parameters
considered in Figs. \ref{ff1}. The results are presented in Figs.
\ref{ff3} in the regime below the threshold where strong
three-photon correlations are realized. As we see, in this case
the correlation function reach to $g^{(3)}=75$ for $n=1.1$ (curves
(1)) for time-intervals between pulses, $\tau = 10\gamma^{-1}$.
Decreasing of the time-separation between pulses leads to
decreasing of the correlation function (see, curves (2), where
$\tau = 5\gamma^{-1}$).

\begin{figure}[h]
\includegraphics[width=15cm]{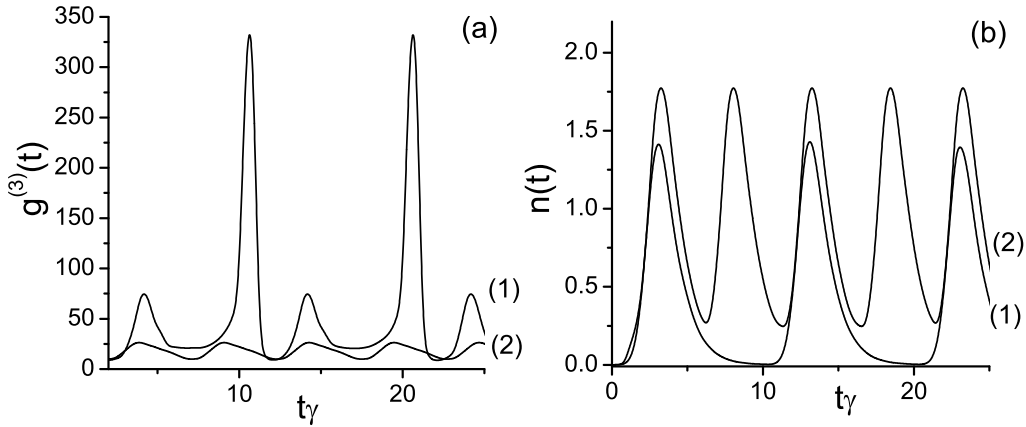}\\
\caption{The correlation functions (a) and the mean photon numbers
(b) at $E_{L}/\overline{E_{th}}=0.2$ for time-separation between
pulses: $\tau \gamma = 10$,  curves (1) ; $\tau \gamma = 5$,
curves (2). The parameters are: $\chi_{1}/\gamma=0.2$,
$\chi_{2}/\gamma=0.1$, $\gamma_{2} = \gamma_{1}=\gamma$, $T \gamma
=1$.}\label{ff3}
\end{figure}

\section{Conclusion}\label{Conclusion}

In conclusion, we have studied quantum properties of photon triplet cardinally different
 from those of twin photons. Because photon-triplet originates from a single laser photon, the quantum correlations
take place between all three photons  allowing the
creation of entangled, non-product states. The production of photon-triplets
in the presence of stimulation radiative processes, cavity
feed-back effects and dissipation have been investigated.  We have demonstrated
the possibility to create
polarization, non-product states of photon-triplet  for
one-passing, collinear  configuration of cascaded parametric spontaneous processes.
We have also  illustrated three-photon structure of sub-harmonic mode on
 the base of both the photon-number distribution and the Wigner function.
We have demonstrated the
operational regimes depending on the durations of pulses and the
intervals between them that guarantees strong three-photon-number
correlations. This effect of strong correlation takes place for
the definite time intervals corresponding to generation of high
intensity "three-photon mode" in over transient regime and for
wide ranges of the system parameters. We hope that these results could be of interest
in areas of quantum communications and photonic quantum computing.

\end{document}